# QuDASH: Quantum-inspired rate adaptation approach for DASH video streaming

Bo Wei[1,2,3], Member, IEEE, Hang Song[4], Member, IEEE, Makoto Nakamura[5], Koichi Kimura[5], Nozomu Togawa[2], Member, IEEE, and Jiro Katto[2], Member, IEEE
[1]Department of Systems Innovation, School of Engineering, The University of Tokyo, Japan
[2]Department of Computer Science and Communication Engineering, Waseda University, Japan
[3]Japan Science and Technology Agency (JST), PRESTO, Kawaguchi, Saitama, Japan
[4]Department of Transdisciplinary Science and Engineering, School of Environment and Society, Tokyo Institute of Technology, Japan
[5]Fujitsu Ltd, Japan

Corresponding author: Hang Song (e-mail: song.h.ac@m.titech.ac.jp)

This work was supported in part by the Fujitsu-Waseda Digital Annealer FWDA Research Project, in part by JSPS KAKENHI Grant No. 20K14740, in part by NICT (Grant No. 03801), and in part by JST, PRESTO (Grant No. JPMJPR22P4), Japan.

**ABSTRACT** Internet traffic is dramatically increasing with the development of network technologies and video streaming traffic accounts for large amount within the total traffic, which reveals the importance to guarantee the quality of content delivery service. Based on the network conditions, adaptive bitrate (ABR) control is utilized as a common technique which can choose the proper bitrate to ensure the video streaming quality. In this paper, new bitrate control method, QuDASH is proposed by taking advantage of the emerging quantum technology. In QuDASH, the adaptive control model is developed using the quadratic unconstrained binary optimization (QUBO), which aims at increasing the average bitrate and decreasing the video rebuffering events to maximize the user quality of experience (QoE). In order to formulate the video control model, first the QUBO terms of different factors are defined regarding video quality, bitrate change, and buffer condition. Then, all the individual QUBO terms are merged to generate an objective function. By minimizing the QUBO objective function, the bitrate choice is determined from the solution. The control model is solved by Digital Annealer, which is a quantum-inspired computing technology. The evaluation of the proposed method is carried out by simulation with the throughput traces obtained in real world under different scenarios and the comparison with other methods is conducted. Experiment results demonstrated that the proposed QuDASH method has better performance in terms of QoE compared with other advanced ABR methods. In 68.2% of the examined cases, QuDASH achieves the highest QoE results, which shows the superiority of the QuDASH over conventional methods.

## I. INTRODUCTION

Video traffic has been the majority of the total global IP traffic, which account for over 80% [1]. In addition, with the deployment of 5th Generation Mobile Communication System (5G), Ultra High-Definition (UHD) video transmission is enabled including 4K/8K video, even virtual reality (VR) and augmented reality (AR) delivery [2, 3]. By 2020, there are already 26 million VR headsets sold and the shipments are expected to reach 43.5 million in 2025 [4]. Furthermore, because of the COVID19 pandemic, the needs for virtual meeting, remote education, remote work have dramatically increased, which leads to the fast growing of video streaming services [5, 6]. In 2022, the video traffic has increased by 24% [7]. These situations solidate the dominance of video traffic in current networks. Therefore, how to ensure the quality of the video streaming is very essential [8].

Since the network is dynamic, when network congestion occurs, the communication speed will drop and cannot support video streaming with high encoded bitrate. If the video is transmitted with constant high bitrate, the video streaming will probably encounter freezing event and the user quality of experience (QoE) will be impaired. To guarantee the user QoE in dynamic networks, video transmission technology named dynamic adaptive streaming over HTTP (DASH) has been proposed and become the de facto standard [9]. In DASH, the video content is encoded in several quality levels with different bitrates. During the video streaming, the user can adaptively select the bitrate level according to various network conditions. If the network is good, high bitrate can be chosen to improve the video quality. Instead, if the network is bad, lower bitrate is preferred to prevent the freezing events. And how to decide the bitrate selection is important in DASH streaming. The adaptive bitrate (ABR) control technique plays an essential role to realize high QoE. Based on the network parameters such as throughput and buffer status, ABR algorithm can select the bitrate level to balance the video quality and video interruption. Thus, the user can enjoy the video streaming with good quality regardless of the communication situation.

The ABR methods in existing literatures can be mainly classified into three kinds: (1) Rate-based [10, 11], (2) Buffer-based [12, 13] and (3) Hybrid [14-16]. Rate-based methods mainly exploit the throughput prediction to decide the future video quality [17-19]. Buffer-based methods mainly consider current buffer status to make the bitrate decision. Hybrid methods take throughput, buffer status, as well as other information into account when determine the chosen bitrate.

Existing methods are mainly based on the classical computers. In contrast, this work makes the initial effort to incorporate the fast-evolving quantum technologies for optimizing the bitrate adaptation in DASH video streaming. In this paper, a novel quantum-inspired ABR method, QuDASH is proposed. It is necessary to formulate the problem as the form which can be solved by quantum technology. In QuDASH, the Ising machine model is utilized and a quadratic unconstrained binary optimization (QUBO) function is derived to obtain the optimal choice of bitrate for video streaming [20]. Generally, the number of potential combinations in QUBO function becomes large with the increase of variables, which makes it difficult to be solved using straightforward solution. Instead, the QUBO problem can be solved efficiently by the mechanism such as quantum annealing. However, quantum annealing is still under development and cannot handle large variables. In this work, a hardware architecture Digital Annealer (DA) [21, 22], is utilized to solve the minimization problem of the QUBO objective function. To the best of our knowledge, our work is the first which employs the quantum-inspired approach to solve the rate adaptation problem in DASH.

To evaluate the performance of the proposed method, experiments are conducted by using simulation with the real world measured throughput. Furthermore, the performance comparison is also carried out with conventional ABR algorithms. Results demonstrate that in 68.2% of the simulation cases, QuDASH has better performance in terms of QoE compared with other methods, indicating the superiority of QuDASH.

The rest of this paper is organized as follows. A brief review of the related work is given in Section II. The proposed QuDASH method is detailed in Section III. In Section IV, the experiment configuration is presented. Section V shows the experiment results and discussion. Finally, the conclusion is given and future work is depicted in Section VI.

## II. RELATED WORK

Many ABR algorithms have been proposed to control the video streaming adaptively according to the network dynamics [23-26]. When the network has large bandwidth, the bitrate with high quality will be chosen. While lower bitrate will be chosen to avoid video interruption when the network is degraded. By properly designing the ABR method, the video quality can be ensured. Based on the control range, the ABR technique includes different approaches. The client-side approach adapts the bitrate selection on the client where only the information obtained from the client application is utilized. While the network-assisted approach introduces the cooperation among the clients, server, and the network to ensure the qualities and fairness of multiple clients [27, 28]. This work focuses on the client-side approach.

Generally, there are three categories of client-side ABR methods in state-of-the-art research, which are the RB methods, BB methods, and hybrid methods.

### 1) RATE-BASED STRATEGY

In RB method, the future throughput is predicted and utilized to decide the bitrate selection. Harmonic mean is a preliminary method among the throughput prediction methods [29, 30]. In the basic RB method, the maximum bitrate which is lower than the predicted throughput value will be chosen for downloading the next video segment. FESTIVE is an advanced method which not only employs the predicted throughput but also utilizes scheduling scheme for bitrate selection [10]. In FESTIVE, a randomized scheduling is set to avoid the bias caused by initial conditions. And the stateful and delayed bitrate update mechanisms were put forward to improve the stability.

### 2) BUFFER-BASED STRATEGY

In BB method, the current buffer occupancy condition is taken into consideration for rate adaptation. Compared with throughput prediction, buffer occupancy is an exact factor. In the basic BB method, the buffer status was mapped to the bitrate selection. When the buffer level is low, lower bitrate will be chosen and vice versa [12]. BOLA is an advanced BB method where the Lyapunov optimization was employed to make rate adaptation [13].

### 3) HYBRID STRATEGY

In hybrid method, not only the throughput prediction and buffer occupancy, but also other information is considered to maximize the user QoE. In [15], a control-theoretic model, MPC was constructed to consider the throughput and buffer occupancy dynamics to decide the bitrate of future chunks. In [14], reinforcement learning based method Pensieve was proposed by constructing the neural network model. In Pensieve, the information of current network is utilized as input, and the bitrate level choices are regarded as output. Then, the neural network was trained using asynchronous advantage actor-critic (A3C) algorithm to figure out the relationship between the input parameters and the bitrate choice.

Existing ABR methods incorporate several emerging techniques such as control theories and machine learning, which are basically solved by classical computers. Compared with the existing works, this work innovatively utilized the quantum-inspired technique for dealing with the video streaming problem. Quantum technology is evolving rapidly nowadays, which is promising to provide much higher computation capability compared with classical computers. Although quantum technology is advanced, the problem needs careful formulation to make it solvable with quantum technology. This work contributes to formulate the video control problem into QUBO functions and propose QuDASH method, which can be solved by the quantum techniques. By taking advantage of the quantum technique, it is expected that the video control problem can be better addressed. This approach is promising to provide a new direction for the future development of video streaming technologies with quantum technique.

## III. QUANTUM-INSPIRED RATE ADAPTATION
In this section, the design and mechanism of QuDASH are presented in detail.

### A. DASH STREAMING AND ABR CONSIDERATION
In the DASH standard, the video content is encoded into different bitrate levels. Meanwhile, the total video content is divided into segments and each segment contains same video duration. During the video streaming, the time for downloading one certain segment is calculated as:

$$\Delta t_n = \frac{S_n(l)}{\int_{t_n}^{t_n+\Delta t_n} C\, dt} \quad (1)$$

where $n$ is the segment index. $l$ is the bitrate level. $S$ is the size of the segment. $C$ is network bandwidth. $t_n$ is the start time of downloading the $n$th segment. In order to make sure the video streaming is resistant to network degradation, the buffer is utilized in video streaming. The buffer occupancy after downloading one certain segment is calculated as follows:

$$B_{n+1} = M - \Delta t_n + B_n \quad (2)$$

where $B_n$ is the buffer size when the $n$th segment starts to be downloaded. $M$ is the video duration of one segment. If $\Delta t_n$ is larger than $B_n$, rebuffering will happen and $B_{n+1}$ becomes $M$ after downloading the segment.

To achieve high-QoE video streaming, several factors are taken into consideration, one of which is the bitrate level. Basically, higher bitrate can contribute to higher QoE. On the contrary, the rebuffering event is more likely to occur when high bitrate is chosen, which will impair the user QoE. Another factor is the variation of video quality. Compared to frequent bitrate change, smooth streaming should be preferred. These mentioned above are the common factors for any ABR algorithm design. In QuDASH, the optimization problem is formulated by reflecting all these factors.

### B. QUDASH DESIGN
QuDASH is a quantum-inspired method designed by employing the Ising machine model. Ising model is originally used to express ferromagnetism, which can be represented as follows:

$$H(\sigma) = -\sum\sum J_{ij}\sigma_i\sigma_j - \sum h_i\sigma_i \quad (3)$$

where $H$ is the Hamiltonian function representing the energy of the configuration. $\sigma$ is the variable depicting the spin state of a site which has only two possible values +1 or -1. $J_{ij}$ is the interaction of two sites and $h_i$ represents the external magnetic field influence. With this model, the phase transition can be analyzed.

Ising model can be related to the QUBO function by simply introducing $\sigma = 2x - 1$ and $x = x^2$ [31]. QUBO function is generally utilized to formulate a combinatorial optimization problem and can be expressed as:

$$f_Q(x) = \sum\sum q_{ij}x_ix_j \quad (4)$$

where $x_i$, $x_j$ are binary variables which can only be assigned as 0 or 1. $q_{ij}$ is the coefficient which determine the QUBO function. The goal of QUBO problem is to find a set of solutions for $x$ which can minimize the QUBO objective function $f_Q$. This kind of problem is difficult to be solved using classical computers because of the large number of variables and potential combinations. However, owing to the close relation to Ising model, QUBO problem can be solved by the emerging annealing technologies such as quantum annealing [32].

To adapt and integrate the quantum techniques for video streaming, the bitrate control problem should be formulated into the QUBO function. The purpose for formulating the QUBO function is that the user QoE can be enhanced by solving this function. Since the user QoE is related to the video quality, bitrate change, and buffer condition, the QUBO function is built considering the above-mentioned factors. In the QUBO function, each variable is binary which only has two values, the bitrate choices need to be represented using several variables and the number is same as the total bitrate levels. Here, assume that the video is encoded in $L$ levels. Then, $L$ binary variable set $\{x_{n,1}, x_{n,2}, …, x_{n,L}\}$ indicate the choice of the $n$th segment. For each segment, only one bitrate should be chosen. Therefore, only the value of one variable in the set can be 1 and others should be 0.

In QuDASH, the video streaming problem is formulated into QUBO objective function considering video quality, bitrate change and rebuffering events. Next, the establishment of QUBO terms relating to different factors are illustrated in detail one by one.

### 1) VIDEO QUALITY FACTOR

In terms of the video quality factor, the purpose is to select as high bitrate as possible. And the QUBO term is defined as:

$$H_{\text{quality}} = -a \sum_{n=1}^{N} \left( \sum_{l=1}^{L} x_{n,l} \cdot q(l) \right) \quad (5)$$

where $q(l)$ indicates the quality metric of the $l$th bitrate level. Since only the variable value for the selected bitrate is 1, the summation of $x_{n,l} \cdot q(l)$ through all possible bitrate choices equals the quality of the selected bitrate. $N$ is the segment number which is considered for optimization. $a$ is the coefficient for this QUBO term. The minus sign is added to the entire term because the optimization of QUBO function is to find the minimum. Through minimizing the QUBO term (5), the larger bitrate level is more likely to be chosen which can make $H_{\text{quality}}$ smaller.

### 2) BITRATE CHANGE FACTOR

In terms of the bitrate change factor, the purpose is to make the video as smooth as possible. And the QUBO term is defined as:

$$H_{\text{change}} = b \sum_{n=1}^{N} \left( \sum_{l=1}^{L} (x_{n,l} - x_{n-1,l}) \cdot q(l) \right)^2 \quad (6)$$

where $b$ is the coefficient for this term. $(x_{n,l} - x_{n-1,l}) \cdot q(l)$ reflects the quality change of the selected bitrates between two consecutive segments. The squared quality change makes the value to be always larger than 0. If there is no change, this is the best case for this term and the value will be 0. On the other hand, if the bitrate changes a lot, the value will be large, which gives more penalty in the QUBO function. Therefore, through minimizing the QUBO term (6), it tends to make the bitrate level unchanged to keep this term small. With $H_{\text{quality}}$, the optimization can guarantee the bitrate change to be kept in a low level, which can benefit the user QoE.

### 3) REBUFFERING EVENT FACTOR

In terms of the rebuffering factor, it is expected that the buffer is kept in the level which will not run out during downloading the next segment to avoid the interruption of the streaming. Therefore, the buffer size should be larger than the downloading time for a certain segment, which can be defined as follows:

$$\frac{\sum_{l=1}^{L} S_n(l) \cdot x_{n,l}}{C_{\text{pred}}} - B_n < 0 \quad (7)$$

where the $S_n(l)$ is the size of $n$th segment at $l$th level. $C_{\text{pred}}$ is the predicted future bandwidth. With this constraint, the optimization problem can suppress the occurrence of rebuffering events. Since QUBO problem cannot include inequality, the slack variables are introduced to transform the inequality into QUBO formulation.

It should be noted that the condition defined in (7) must be satisfied for each segment. Therefore, there are totally $N$ inequalities to be considered in optimization. Considering the relationship between the buffer sizes of two consecutive segments, the inequalities can be re-formulated as:

$$\begin{cases} \dfrac{\sum_{l=1}^{L} S_1(l) \cdot x_{1,l}}{C_{\text{pred}}} < B_1 \\ \dfrac{\sum_{l=1}^{L} S_2(l) \cdot x_{2,l}}{C_{\text{pred}}} < B_1 - \dfrac{\sum_{l=1}^{L} S_1(l) \cdot x_{1,l}}{C_{\text{pred}}} + M \quad (8) \\ \quad\quad\quad\quad\quad \vdots \\ \dfrac{\sum_{l=1}^{L} S_N(l) \cdot x_{N,l}}{C_{\text{pred}}} < B_1 - \sum_{n=1}^{N-1} \dfrac{\sum_{l=1}^{L} S_n(l) \cdot x_{n,l}}{C_{\text{pred}}} + (N-1)M \end{cases}$$

where only the variables of the bitrate selection are contained. $B_1$ is the buffer occupancy size before the downloading of the next segment. The general formula of (8) for the $n$th segment can be written as:

$$\sum_{i=1}^{n} \frac{\sum_{l=1}^{L} S_i(l) \cdot x_{i,l}}{C_{\text{pred}}} < B_1 + (i-1)M \quad (9)$$

Let $U = B_1 + (i-1)M$ and $w_{i,l} = S_i(l)/C_{\text{pred}}$, and (9) can be expressed as:

$$\sum_{i=1}^{n} \sum_{l=1}^{L} w_{i,l} \cdot x_{i,l} < U \quad (10)$$

To satisfy the constraint by (10) with QUBO term, slack variables $y_k$ are introduced and the QUBO term is defined as:

$$H_n = \left( \sum_{k=0}^{K-1} 2^k \cdot y_k - 2^K + 1 + U - \sum_{i=1}^{n} \sum_{l=1}^{L} w_{i,l} \cdot x_{i,l} \right)^2 \quad (11)$$

where $K$ is the number of introduced slack variables. $K$ is determined as the smallest integer which is larger than $\log_2 U$. Since $y_k$ is binary variable, $\sum_{k=0}^{K-1} 2^k \cdot y_k - (2^K - 1)$ is non-positive. On the other hand, $w_{i,l}$ is positive value and $\sum_{i=1}^{n} \sum_{l=1}^{L} w_{i,l} \cdot x_{i,l}$ is non-negative. Therefore, there is solution which can minimize (11). And when (11) is minimized, it can ensure that (10) is also satisfied.

### 4) ONE BITRATE CONSTRAINT

In DASH streaming, only one bitrate can be chosen for each segment. Since the variable in QUBO can be only assigned with 1 or 0, the bitrate choice of one segment is modelled as several variables where each variable shows whether a certain bitrate level is chosen. For the future segment, the summation of these variables should be 1 to make sure that only one bitrate level is chosen. This constraint is satisfied by using the following QUBO term:

$$H_{\text{one}} = \sum_{n=1}^{N}\left(\sum_{l=1}^{L} x_{n,l} - 1\right)^2 \quad (12)$$

Since $x_{n,l}$ is binary variable, when $\sum_{l=1}^{L} x_{n,l} = 1$, only one variable is 1. At the same time, (12) can reach the minimum value. Therefore, (12) ensures that only one bitrate is chosen.

### 5) QUBO OBJECTIVE FUNCTION

Finally, after establishing the four individual QUBO terms, the QUBO objective function for DASH streaming optimization is formed by merging the individual terms as follows.

$$\begin{aligned} H = &-a \sum_{n=1}^{N}\left(\sum_{l=1}^{L} x_{n,l} \cdot q(l)\right) \\ &+ b \sum_{n=1}^{N}\left(\sum_{l=1}^{L}(x_{n,l} - x_{n-1,l}) \cdot q(l)\right)^2 \\ &+ c \sum_{n=1}^{N}\left(\sum_{l=1}^{L} x_{n,l} - 1\right)^2 \\ &+ d \sum_{n=1}^{N}\left(\sum_{k=0}^{K-1} 2^k \cdot y_k - 2^K + 1 + U - \sum_{i=1}^{n}\sum_{l=1}^{L} w_{i,l} \cdot x_{i,l}\right)^2 \end{aligned} \quad (13)$$

where *a*, *b*, *c*, *d* are the coefficients for each term, respectively. By solving the minimization problem, the optimal solution can be obtained as:

$$\{x_{i,l}\}_{\text{opt}} = \arg\min H(x_{i,l}) \quad (14)$$

which represents the selection of the bitrate level. By checking which variable is 1 in the solution, the corresponding bitrate level will be chosen for the next segment downloading.

## IV. EXPERIMENT CONFIGURATION
### A. DASH STREAMING AND ABR CONSIDERATION

In order to evaluate the performance of the proposed QuDASH method and compare with existing ABR methods, simulation experiments were conducted.

The video content used for evaluation is BigBuckBunny [33]. It is encoded from the source file to 6 bitrate levels, which are {40, 16, 8, 5, 2.5, 1} Mbps. The resolutions of these bitrate are set as {2160, 1440, 1080, 720, 480, 360} p. The highest resolution is 4K in the video transmission. For DASH streaming, the total content is divided into video chunks with video duration of 2 seconds.

In the experiments, the network trace data measured in real world was utilized. These data were collected by self-developed Android application, and the downloading of data from the lab server were executed through the mobile networks in different scenarios. It is assumed that user is enjoying video streaming when measuring the throughput data. In the measurement, one user device was connected with the server through LTE provided by the major public cellular carrier. The mobile device measured and recorded throughput by downloading the video segment per second from content server. Totally, three network scenarios were utilized for evaluation including static, walking, and on a moving bus. The network dynamics of these scenarios are different. The measured traces are shown in Fig. 1.

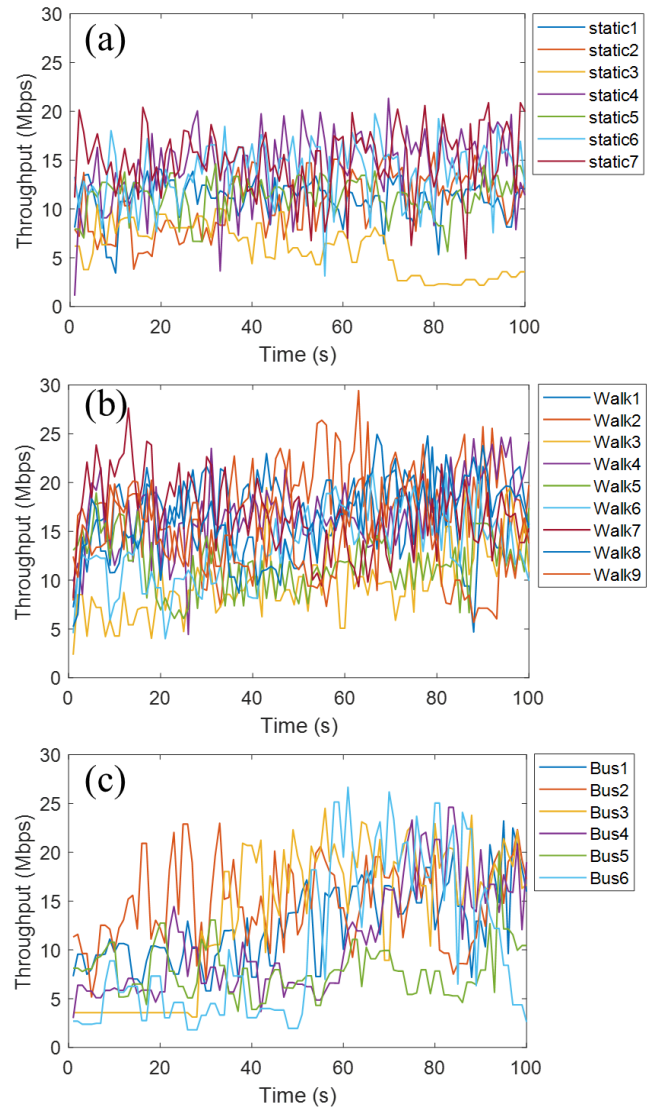

**FIGURE 1.** Network traces used in the experiment. (a) Static cases. (b) Walk cases. (C) Bus cases.

The dataset of static scenario contains seven throughput traces, each of which was collected at the 1 second granularity in the lab or at home when the user was static at different times. For each trace, the duration of 100-second data is utilized. We can observe that the throughput is stationary in each trace, the value of which fluctuate very small with the change of time.

As to the dataset of walking scenario, it was collected when user was walking on the street between the lab and home at different times. The dataset of walking user condition is shown in Fig. 1(b), there is fluctuation to some extent compared with the static cases. The throughput traces were also recorded at the 1 second granularity and nine traces were utilized in the experiment.

As to the dataset of moving bus scenario, the user collected the data when riding on campus bus between the Nishi-waseda campus and Waseda campus of Waseda Univeristy at different times. The throughput condition of user on a moving bus is shown in Fig. 1(c), six throughput traces were recorded at 1 second granularity. It can be observed that the throughput of bus condition fluctuates larger than the static and walk case. This is because bus was moving with a larger speed and the communication condition changed severely.

### B. EVALUATION METRICS

In this work, the QoE metric inherited from [14, 15] is utilized to quantitatively evaluate the proposed method. The same metric is also applied to other conventional methods for comparison. The QoE is expressed as:

$$QoE = \sum_{n=1}^{N} q[l_n] - w \sum_{n=1}^{N} T_{\text{rebuf}} - \sum_{n=1}^{N} |q[l_n] - q[l_{n-1}]| \quad (15)$$

where $q[l_n]$ is the quality of bitrate level for the $n$th segment. $T_{\text{rebuf}}$ is the encountered rebuffering time when downloading a certain segment. $w$ is the penalty coefficient applied to the rebuffering term. In this work, $w$ is set as 40 which is the highest bitrate level. The last term indicates the penalty of quality change between two consecutive segments. In the posterior analysis, the average QoE per chunk is utilized for performance comparison. It is calculated by using the total QoE metric divided by the number of chunks in the video.

### C. SOLVING QUDASH WITH DA

For conventional computers, there is difficulty to solve the combinatorial optimization problem since there are extremely large number of potential combinations with the increase of variables. To efficiently solve such Ising model or QUBO problems, there are many proposed mechanisms including classic annealers, dynamic system solvers, and quantum annealing [34]. Quantum annealing is a promising technology, but it is currently still under development and is hard to achieve high-dimensional stable application. In this work, a quantum-inspired architecture named DA provided by FUJITSU is utilized to solve the QuDASH problem [21, 22].

In DA, the simulated annealing with Markov chain Monte Carlo (MCMC) method is utilized and the acceleration is realized by using digital-CMOS-circuits and parallel structure. In the solving process, an energy function is defined as:

$$E(x) = -\sum_{i=1}^{N_v} \sum_{j=1}^{N_v} W_{ij} x_i x_j - \sum_{i=1}^{N_v} b_i x_i \quad (16)$$

where $N_v$ is the total binary variable number in the final objective function. $W_{ij}$ are the connection weight between two different variables. $b_i$ is the bias term. Note that (13) can be reformulated into the form in (16) by introducing $x^2 = x$. At the beginning, an arbitrary initial state is set as $x_0 = \{x_1, x_2, ..., x_{N_v}\}$. Next, one of the variable $x_k$ is proposed to be flipped to 1-$x_k$. Then, the energy between the proposed state $x'$ and the former state $x$ is calculated as:

$$\Delta E(x') = E(x') - E(x) \quad (17)$$

Further, the acceptance probability is calculated using the Metropolis-Hastings algorithm as:

$$A(x'|x) = \min(1, e^{-\beta \Delta E(x')}) \quad (18)$$

where $\beta$ is the inverse of temperature $T$. If $A(x'|x)$ is larger than a random number $u$ with the uniform distribution between 0 and 1, the proposed state update can be accepted. Otherwise, the state is not changed.

When the solving process is carried out in a single thread, the efficiency is low since there is probability that the proposed update is rejected. To accelerate the solution, DA is equipped with the parallel mechanism. For each trial, numerous neighbor states can be examined at the same time. Then the state will be updated by synthetically considering the total results. With the parallel mechanism, the computation time for the QUBO problem is reduced significantly. In DA, the number of the parallel processing thread can be designated.

During the video streaming, when downloading of the previous segment is completed, the objective function of (13) will be updated based on the current streaming status. Then, (13) will be rewritten as (16) which is an equivalent form of the QUBO objective function. Next, the DA is applied to minimize (16). The coefficients $W_{ij}$ and $b_i$ of the equivalent form (16) are extracted and utilized as input to define the problem for the DA. Then, the DA will conduct the optimization and produce the solution. According to the solution, the bitrate choice of the next segment can be determined.

### V. EXPERIMENT RESULTS

Experiments were conducted in various network conditions to investigate the performance of the proposed QuDASH. The network traces were LTE data in different scenarios when user was keeping static, walking and riding on bus.

First, the influence of the parameters $a$, $b$, $c$, $d$, in the objective function (13) is explored. Meanwhile, the parameters $N_{\text{run}}$ and $N_{\text{ite}}$ in DA for solving the optimization problem are also studied. Under different network conditions, these parameters are set in the value range as shown in TABLE I. For the parameters $a$, $b$, $d$, the value is changed from 1 to $10^4$, and the value for $c$ is changed from $10^2$ to $10^6$. The purpose is to explore the best parameter setting for the QuDASH model. $N_{run}$ is defined as the number of annealing processes, which specifies the parallel annealing process threads. For the parameter exploration of $N_{run}$, the value is set as 32, 64, and 128. Larger value means larger number of parallel annealing processes. $N_{ite}$ represents the number of iterations for solving the QUBO problem of one annealing process.

TABLE I
PARAMETER SETTING

| Parameter | Value | Parameter | value |
|---|---|---|---|
| $a$ | $1 \sim 10^4$ | $d$ | $1 \sim 10^4$ |
| $b$ | $1 \sim 10^4$ | $N_{run}$ | 32, 64, 128 |
| $c$ | $10^2 \sim 10^6$ | $N_{ite}$ | $10^2 \sim 10^7$ |

### A. EVALUATION OF QUDASH IN STATIC SCENARIO

To evaluate the performance of QuDASH and the parameter influence, the obtained QoE values under different conditions are analyzed by exploring the parameters individually.

In the static user case, six experiments were carried out for each trace to investigate $N_{ite}$. As for the parameter setting, $a$, $b$, $c$, $d$, $N_{run}$ were fixed as $10^3$, 1, $10^6$, 1, 128, respectively. $N_{ite}$ was changed from $10^2$ to $10^7$ logarithmically. In Fig. 2. It can be observed that when $N_{ite}$ decreases, QoE generally decreases in all the experiments. As shown in Fig. 2, QoE is the smallest in majority of the trials when $N_{ite}$ is set as $10^4$. This is because when the iteration is small, the optimized solution is less likely to achieve global optimization. Contrarily, when $N_{ite}$ is set as large as $10^7$, it is more likely to achieve the global optimization.

In the next experiment, the parameter settings were as follows. $a$, $b$, $c$, $d$, $N_{ite}$ were fixed as $10^3$, 1, $10^6$, 1, $10^7$, respectively. And $N_{run}$ was changed as 32, 64, 128. For each trace, the experiments were conducted 3 times by changing $N_{run}$. From Fig. 3, it can be observed that QoE does not change too much with the change of $N_{run}$. But there is a slight tendency that except for the case of Static1, QoE may decrease when $N_{run}$ is smaller. Therefore, more parallel annealing processes are able to help find the optimal solution of QuDASH.

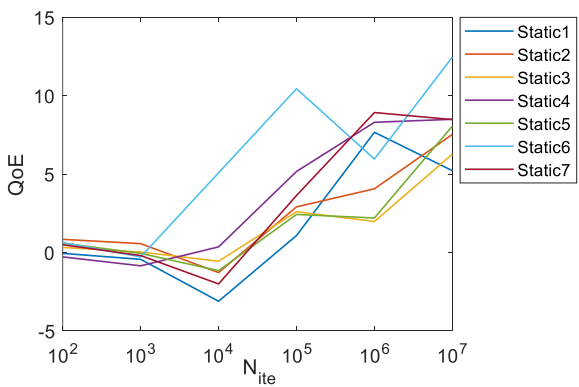

**FIGURE 2.** QoE results of QuDASH with the change of $N_{ite}$ in static case.

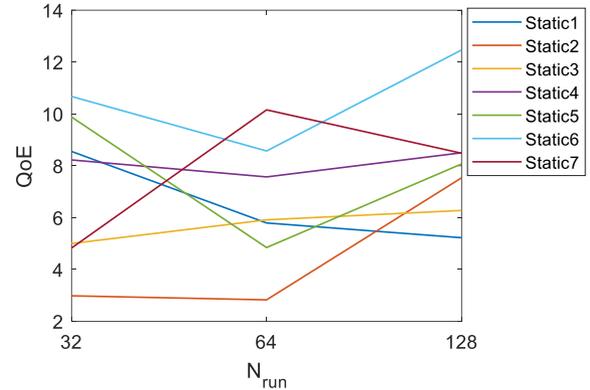

**FIGURE 3.** QoE results of QuDASH with the change of $N_{run}$ in static case.

Then, the influence of parameters in the objective function were investigated. $b$, $c$, $d$, $N_{run}$, $N_{ite}$ were fixed as 1, $10^6$, 1, 128, $10^7$, respectively. And $a$ was varied from 1 to $10^4$. For each trace, the experiments were conducted 5 times by changing $a$. From Fig. 4, it can be observed that QoE of most trials tend to increase. This is because with larger value of $a$, the term for the video quality in objective function is assigned with larger weight, thus it tends to keep the video quality with large value during optimization. In the static user condition, as bandwidth is relatively stable, future network condition evolution tends to not change drastically. Therefore, the video streaming is smooth and the quality factor is the major part which affects QoE during optimization.

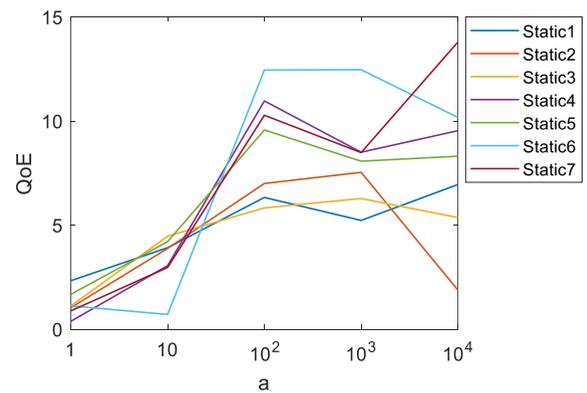

**FIGURE 4.** QoE results of QuDASH with the change of $a$ in static case.

In the next experiment, parameter $b$ was investigated. The parameter settings were as follows. $a$, $c$, $d$, $N_{run}$, $N_{ite}$ were fixed as $10^3$, $10^6$, 1, 128, $10^7$, respectively. And $b$ was varied from 1 to $10^4$. For each trace, the experiments were conducted 5 times by changing $b$. From Fig. 5, it can be observed that with the increasing of $b$, the QoE in most cases is initially increasing and then decreasing. When $b$ was set as $10^4$, QoE is the worst. It can be observed that the strict requirement of less bitrate change will impair the video streaming. Therefore, to ensure overall QoE, the weight for the term of bitrate change in the objective function should not be set too large. $b$ is better to be assigned to the value from 1 to $10^2$.

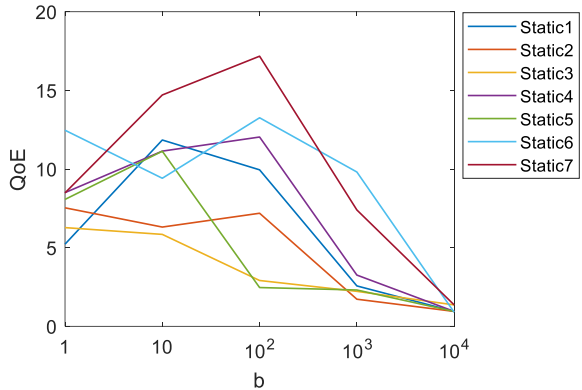

**FIGURE 5. QoE results of QuDASH with the change of $b$ in static case.**

Similarly, the parameters of $c$ and $d$ were also investigated. When investigating $c$, the parameter settings for $a$, $b$, $d$, $N_{run}$, $N_{ite}$ were $10^2$, $10^2$, 1, 128, $10^7$, respectively. While when investigating $d$, the parameter setting for $a$, $b$, $c$, $N_{run}$, $N_{ite}$ were $10^2$, $10^2$, $10^6$, 128, $10^7$. The QoE results with the change of $c$ and $d$ are shown in Fig. 6 and Fig. 7. From Fig. 6, it is difficult to identify the influence of $c$ since the tendency is not obvious. However, theoretically $c$ should be larger to make sure only one bitrate is selected which is the basic requirement in DASH streaming. From Fig. 7, it can be observed that there is a general tendency that QoE is decreasing with $d$. This implies that if the inequality for buffer is strictly satisfied, the QoE may be impaired. Since the prediction of bandwidth may be somehow erroneous, the term for buffer requirement shows variation. Therefore, this term should not be assigned to large weight.

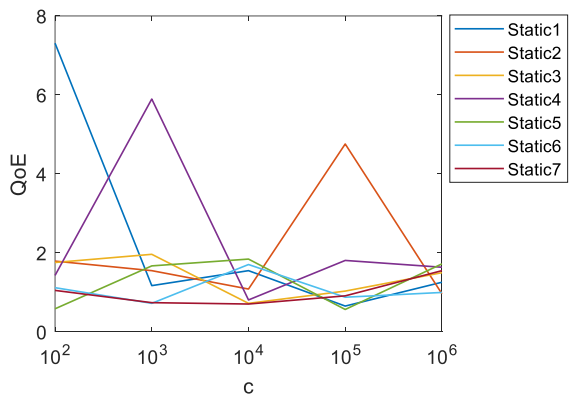

**FIGURE 6. QoE results of QuDASH with the change of $c$ in static case.**

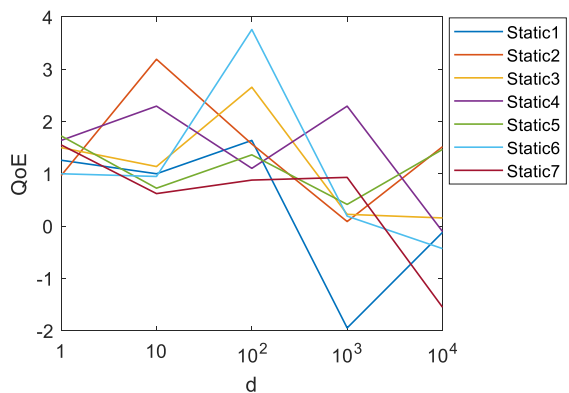

**FIGURE 7. QoE results of QuDASH with the change of $d$ in static case.**

### B. EVALUATION OF QUDASH IN WALK SCENARIO

Similar as static scenarios, the evaluation of QuDASH and influence of the parameters on QoE results for walk scenarios were also investigated. The parameter settings are summarized in TABLE II. The value denoted as $V$ is the varying parameter, which can be referred to in the last column of TABLE II.

TABLE II
PARAMETER SETTINGS IN WALK CASES

| Experiment round | $a$ | $b$ | $c$ | $d$ | $N_{run}$ | $N_{ite}$ | Varying Parameter $V$ |
|---|---|---|---|---|---|---|---|
| 1 ($N_{ite}$) | $10^3$ | 1 | $10^6$ | 1 | 128 | $V$ | $10^2 \rightarrow 10^7$ |
| 2 ($N_{run}$) | $10^3$ | 1 | $10^6$ | 1 | $V$ | $10^7$ | 32, 64, 128 |
| 3 ($a$) | $V$ | 1 | $10^6$ | 1 | 128 | $10^7$ | $1 \rightarrow 10^4$ |
| 4 ($b$) | $10^3$ | $V$ | $10^6$ | 1 | 128 | $10^7$ | $1 \rightarrow 10^4$ |
| 5 ($c$) | $10^2$ | $10^2$ | $V$ | 1 | 128 | $10^7$ | $10^2 \rightarrow 10^6$ |
| 6 ($d$) | $10^2$ | $10^2$ | $10^6$ | $V$ | 128 | $10^7$ | $1 \rightarrow 10^4$ |

In the first experiment, $N_{ite}$ was changed from $10^2$ to $10^7$. Totally six trials were utilized and evaluated. From Fig. 8, it can be observed that QoE decreases when the iteration number is decreasing. This result is consistent with the static case. As for $N_{run}$ which was changed from 32 to 128, the QoE results are shown in Fig. 9. It can be observed that the results are not changed too much, which is also similar to the static scenarios.

The influence of parameters in the objective functions were also investigated. The QoE results when $a$ was changed from 1 to $10^4$ are shown in Fig. 10. It can be observed that the QoE of most cases tend to increase when the value of $a$ increases. The tendency is consistent with the static scenarios. The QoE results when $b$ was changed from 1 to $10^4$ are shown in Fig. 11. It can be observed that initially QoE is almost increasing and then decreasing when $b$ increases. If $b$ is set as large value, QoE will decrease. The best setting for $b$ should be lower than $10^2$. This observation is also similar with static scenarios. As for $c$ and $d$, the QoE results are shown in Fig. 12 and Fig. 13, respectively. From Fig. 12, it can be obtained that the influence of $c$ is random and no obvious trend is recognized. From Fig. 13, it can be noted that when $d$ is set as large value, the QoE will decrease significantly. Although the influence of $d$ is not obvious , $d$ should be set smaller. These observations in walk case are basically the same as in the static cases.

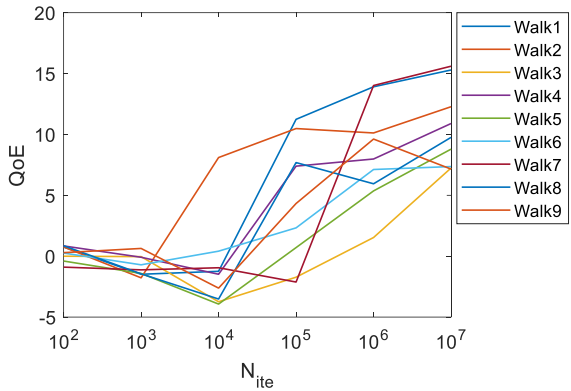

**FIGURE 8.** QoE results of QuDASH with the change of $N_{ite}$ in walk case.

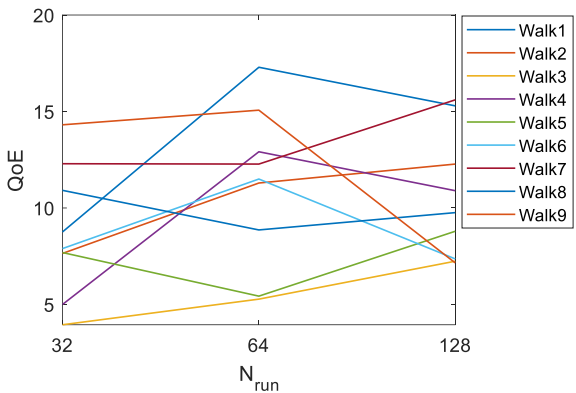

**FIGURE 9.** QoE results of QuDASH with the change of $N_{run}$ in walk case.

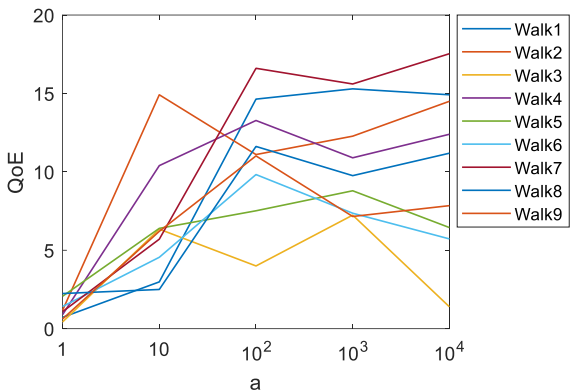

**FIGURE 10.** QoE results of QuDASH with the change of $a$ in walk case.

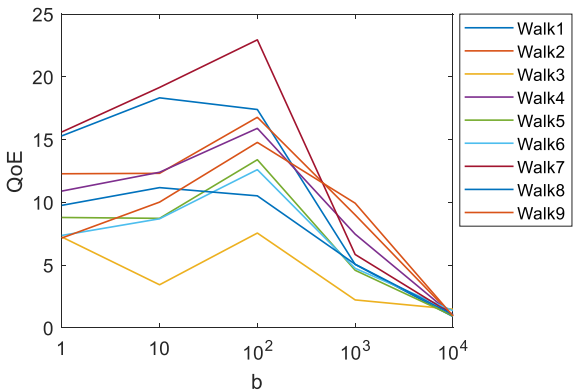

**FIGURE 11.** QoE results of QuDASH with the change of $b$ in walk case.

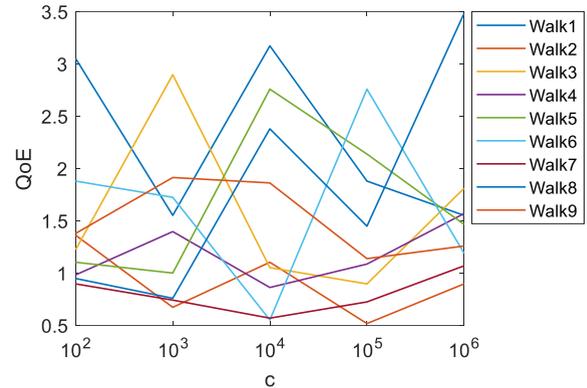

**FIGURE 12.** QoE results of QuDASH with the change of $c$ in walk case.

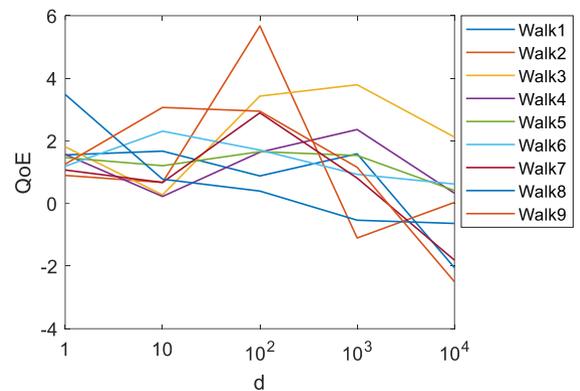

**FIGURE 13.** QoE results of QuDASH with the change of $d$ in walk case.

### C. EVALUATION OF QUDASH IN BUS SCENARIO

For the cases when user is on the moving bus, evaluation was carried out similarly as the former two scenarios. The parameter setting in TABLE II were utilized. The QoE results of different experiments are summarized and shown in Fig. 14. As for $N_{ite}$ shown in Fig. 14(a), similar trend can be observed as the former two scenarios. Therefore, it can be concluded that the iteration number should be set as large value to obtain better result. As for $N_{run}$ shown in Fig. 14(b), it is difficult to conclude the QoE trend with the change of $N_{run}$. However, considering about efficiency, $N_{run}$ is better to be set as larger value. As to parameter $a$ shown in Fig. 14(c), the increasing trend can be observed with the increase of $a$. This observation is similar to the former two scenarios. Therefore, it can be concluded that $a$ should be set as large value to obtain better QoE result. As for $b$ shown in Fig. 14(d), it can be observed that when $b$ is set as large value, QoE is impaired. From experiment results of the former two scenarios, it can be concluded that $b$ should be given small value. As for $c$ shown in Fig. 14(e), no strong relation is found between $c$ and QoE. However, $c$ should be large value to ensure only one bitrate is chosen, which is also the same in the other two scenarios. As for $d$, it can be observed in Fig. 14(f) that QoE decreases with the increasing of $d$. Therefore, $d$ should be set as small value.

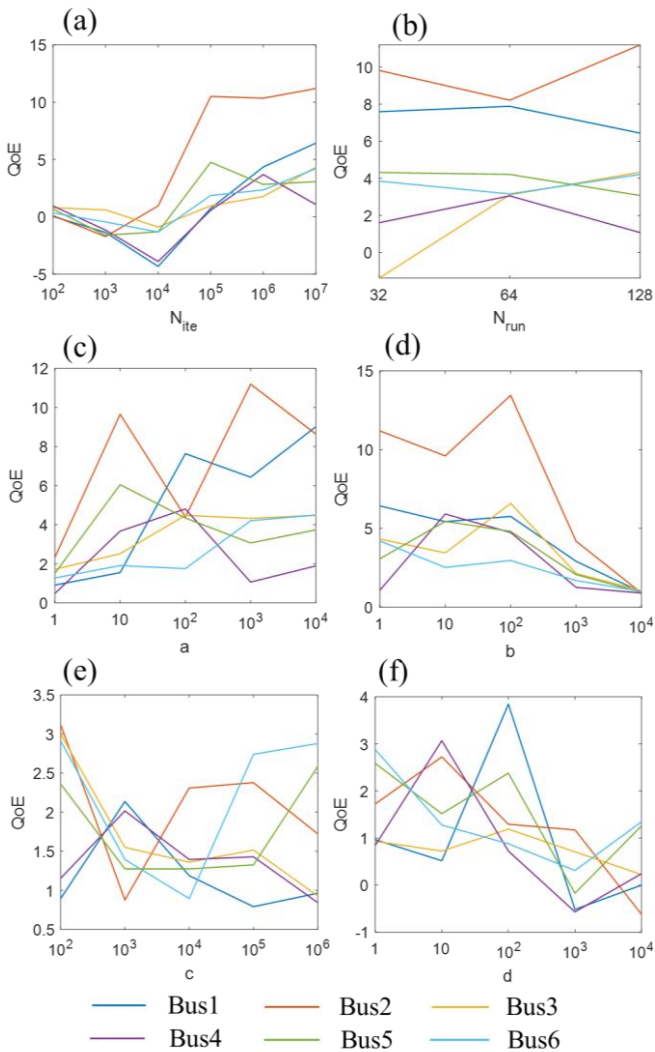

**FIGURE 14.** QoE results with different parameter setting of QuDASH in bus case. (a) Change of $N_{ite}$. (b) Change of $N_{run}$. (c) Change of $a$. (d) Change of $b$. (e) Change of $c$. (f) Change of $d$.

### D. COMPARISON WITH CONVENTIONAL METHODS

To evaluate the superiority of proposed QuDASH method, the comparison with other conventional methods is conducted, including RB, BB [12], MPC [15], and reinforcement learning (RL) [14] methods. These methods are chosen since they represent a wide spectrum of the existing ABR categories. In RB method, the harmonic mean of the throughputs when downloading previous 5 segments is calculated as the prediction of the future throughput. Then, the bitrate level which is lower than the predicted throughput is chosen for the next video segment. In BB method, the reservoir and cushion are set as 5 seconds and 55 seconds, respectively. In MPC, buffer state and throughput prediction with harmonic mean is utilized and the future 5 segments are considered for maximization of QoE. In RL, the setting is kept the same as the default in [14]. For all the methods, the maximum buffer limit is set as 60 seconds. The experiments were conducted in all the three scenarios and the QoE were calculated and compared.

For the static scenario, QoE results of different methods are shown in Fig. 15. It can be observed that in 71.4% cases, QuDASH shows the best QoE. By averaging through all cases, the average QoE of QuDASH is the highest, which is 9.8% higher than the second-best method. Compared with other methods, QuDASH shows much better performance. The maximum difference of average QoE between QuDASH and conventional methods is as large as 8.4, and QoE of QuDASH is 3.6 times of the BB method.

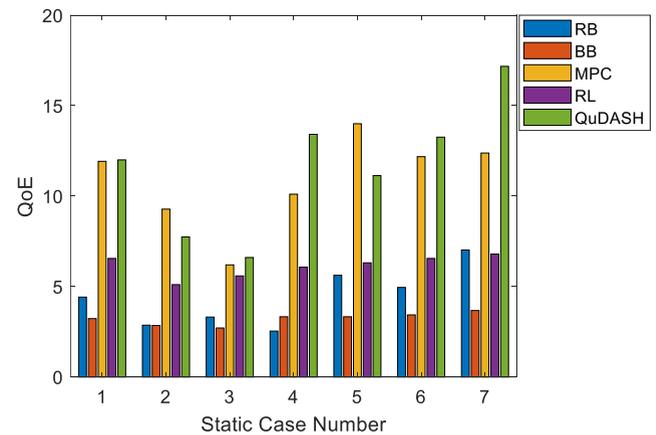

**FIGURE 15.** Comparison of QoE results by different methods in static scenario.

For the walk scenario, QoE results of different methods are shown in Fig. 16. In this scenario, QuDASH also performs the best and the QoE of QuDASH is the highest in 77.8% cases. In terms of average QoE, QuDASH also achieves the highest, which is 12.7% higher than the second-best method. Meanwhile, QuDASH also shows much better performance that the maximum difference of average QoE between QuDASH and other methods is as large as 11.5, and QoE of QuDASH is 4.3 times of the BB method.

For the bus scenario, QoE results of different methods are shown in Fig. 17. In this scenario, QuDASH shows the best performance in half of the cases. In terms of the average QoE throughout all cases, QuDASH is 3% lower than MPC method. Although QuDASH is not the best in bus scenario, it shows competitive performance with MPC method. Besides, QuDASH still holds significant advantage compared with the other methods except for MPC.

Considering all the scenarios together, the cumulative distribution function (CDF) plot of QoE results for different methods are shown in Fig. 18. It can be observed that, although QuDASH does not outperform conventional ABR methods in all the cases, it achieves the best performance in 68.2% cases. From the viewpoint of statistics, it can be concluded that the proposed QuDASH has superiority than existing methods.

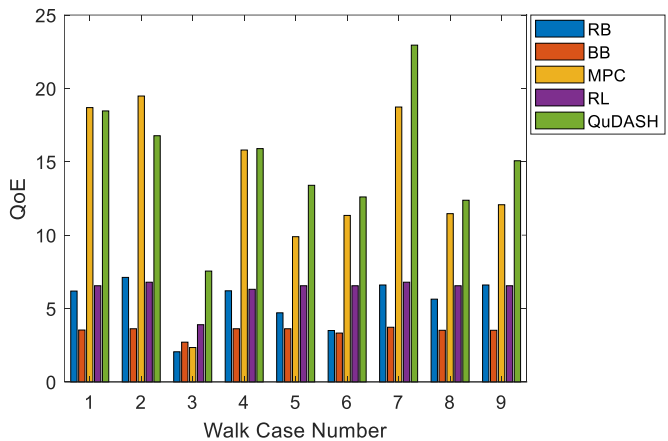

**FIGURE 16.** Comparison of the QoE results by different methods in walk scenario.

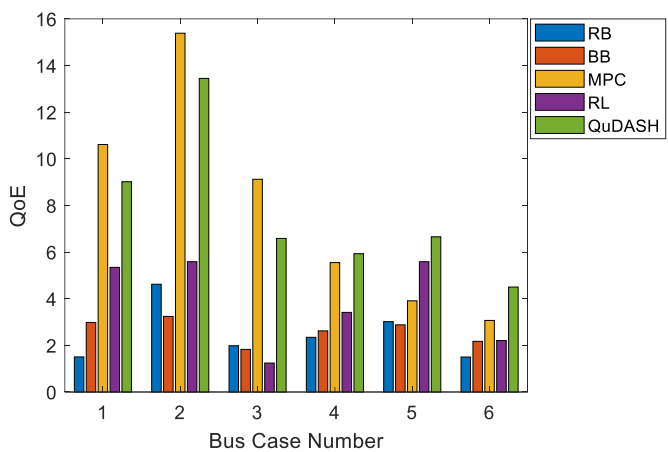

**FIGURE 17.** Comparison of QoE results by different methods in bus scenario.

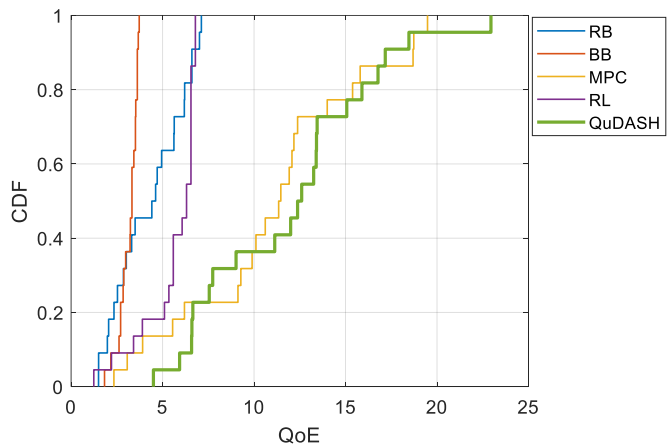

**FIGURE 18.** The CDF plot of the QoE results by different ABR methods.

## VI. CONCLUSION

In this paper, a quantum-inspired ABR approach QuDASH was proposed by incorporating the emerging quantum technology while the existing methods are basically developed and solved by classical computers. In this work, QuDASH was proposed by formulating the adaptive video streaming problem as Ising/QUBO model and solved by quantum-inspired architecture, DA, which is a digital-CMOS hardware developed to accelerate the solving process of combinatorial problem. By conducting experiments with real measured network traces in three different scenarios, the performance of QuDASH was evaluated. Experiment results demonstrated that the QoE of QuDASH are the best in most cases compared with conventional methods, which shows the superiority and potentials of proposed QuDASH method. This work is promising to open a new technique category for video streaming with quantum technologies. In the future, the proposed method will be evaluated in field test under various scenarios and the QUBO model in QuDASH will be further optimized. Meanwhile, the comparison with more state-of-the-art methods will be carried out.

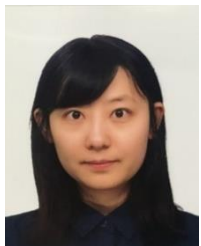
**BO WEI** received the B.E. and M.E. degrees from Tianjin University, Tianjin, China, in 2012 and 2015, respectively. She received the Ph.D. degree from Waseda University, Tokyo, Japan in 2019. From 2019 to 2023, she was an assistant professor with the Graduate School of Fundamental Science and Engineering, Waseda University. She is currently a specially appointed assistant professor with the University of Tokyo and the PRESTO researcher with JST.

Her research interests include wireless communication, machine learning, adaptive video transmission, computer networking, quantum computing and Internet of Things.

She is a member of the IEEE, ACM and IEICE.

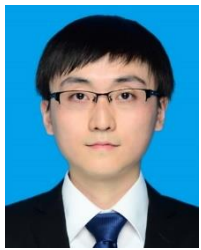
**HANG SONG** received the B.S. and M.S. degrees in electronic science and technology from Tianjin University, Tianjin, China, in 2012 and 2015, respectively. He received the Ph.D. degree from Hiroshima University, Hiroshima, Japan in 2018. From 2019 to 2020, he was a lecturer with the school of microelectronics, Tianjin University, China. From 2020 to 2022, he was a specially appointed researcher with the University of Tokyo, Japan. He is currently an assistant professor with Tokyo Institute of Technology. He is also a visiting researcher of the University of Tokyo and a visiting lecturer of Hiroshima University.

His research interests are wireless sensing, information and communication networks, microwave detection and imaging, biomedical engineering.

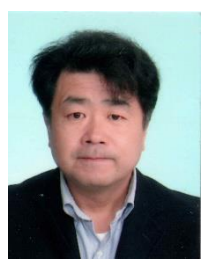
**MAKOTO NAKAMURA** received the B.E. degrees from Shimane University, Matsue, in 1985. After that, he joined Fujitsu Ltd. and has been developing LSI processes and relative materials. Based on these research results, he received a Ph.D. degree from Shizuoka University, Hamamatsu, Japan, in 2006. Currently, he is exploring new applications for new computers (Digital Annealers) developed by Fujitsu Ltd., inspired by quantum annealers. He has also recently begun exploring applications for quantum computing.

He is a member of the Japan Society of Applied Physics and the Japan Society of Vacuum and Surface Science.

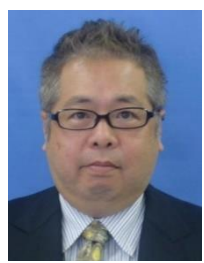
**KOICHI KIMURA** received the B.E. degrees from Shinshu University, nagano, in 1990. After that, he joined Fujitsu Ltd. and has been developing Lightweight material and bioplastic for PC's housing. Currently, he is exploring new applications for new computers (Digital Annealers) developed by Fujitsu Ltd., inspired by quantum computing. He has also recently begun researchng applications for quantum computing.

He is a member of Japan Society of Mechanical Engineers

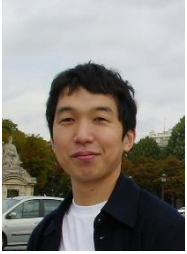

**NOZOMU TOGAWA** received the B. Eng., M. Eng., and Dr. Eng. degrees from Waseda University in 1992, 1994, and 1997, respectively, all in electrical engineering. He is presently a Professor in the Department of Computer Science and Communications Engineering, Waseda University. His research interests are VLSI design, graph theory, and computational geometry. He is a member of IEICE and IPSJ.

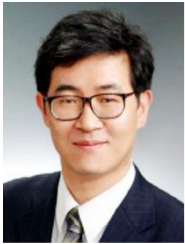

**JIRO KATTO** received the B.S., M.E., and Ph.D. degrees from the University of Tokyo, Tokyo, Japan, in 1987, 1989, and 1992, respectively. He was with NEC Corporation from 1992 to 1999. He was a Visiting Scholar with Princeton University, NJ, USA, from 1996 to 1997. He then joined Waseda University in 1999, where he is currently a professor. His research interest include multimedia signal processing and multimedia communication systems.

He is a Fellow of IEICE and a member of ACM.